\documentclass[aps,prl,twocolumn,floatfix,showpacs]{revtex4-1}
\usepackage{amsmath,amssymb,graphicx,bm} 

\begin{document} \title{Continuous RSB mean-field solution of the
  Potts glass}

\author{V.  Jani\v{s}} \author{A. Kl\'\i\v{c}}

\affiliation{Institute of Physics, Academy of Sciences of the Czech
  Republic, Na Slovance 2, CZ-18221 Praha, Czech Republic }
\email{janis@fzu.cz}

\date{\today}


\begin{abstract} We investigate the p-state mean-field model of the
  Potts glass ($2\le p \le 4$) below the continuous phase transition
  to a glassy phase. We find that apart from a solution with a first
  hierarchical level of replica-symmetry breaking (1RSB), locally
  stable close to the transition point, there is a continuous full
  replica-symmetry breaking (FRSB) solution. The latter is marginally
  stable and has a higher free energy than the former. We argue that
  the true equilibrium is reached only by FRSB, being globally
  thermodynamically homogeneous, whereas 1RSB is only locally
  homogeneous. 
  \end{abstract} 

\maketitle 

The Potts model is a generalization of the Ising model to more than
two spin components. It has been a subject to intense research in
recent decades. The original formulation of Potts \cite{Potts52} with
a Hamiltonian $H_{p}=-\sum_{i<j}J_{ij}\delta_{n_{i},n_{j}}$ where
$n_{i}=0,\ldots, p-1$ is an admissible value of the $p$-component model
on the lattice site $\mathbf{R}_{i}$, is unsuitable for practical
calculations. The Potts Hamiltonian can be represented via interacting
spins \cite{Wu82} %
\begin{equation}\label{eq:Hamiltonian}
  H_{p}=-\frac{1}{2}\sum_{i,j}J_{ij}\mathbf{S}_{i}\cdot \mathbf{S}_{j}\ ,
\end{equation} %
where $\mathbf{S}_{i} = \{s_{i}^{1},\ldots s_{i}^{p-1} \}$ are Potts
vector variables taking values from a set of state vectors
$\{\mathbf{e}_{A}\}_{A=1}^{p}$. Functions on vectors $\mathbf{e}_{A}$
are in equilibrium fully defined through their scalar product
$\mathbf{e}_{A}\cdot\mathbf{e}_{B} \equiv
e^{\alpha}_{A}e^{\alpha}_{B}=p \delta_{A B}-1$,
$\alpha\in\{1,...,p-1\}$. We use the Einstein summation convention for
repeating Greek indices of the vector components indicating a scalar product of Potts vectors.

Interest in the Potts model was recently revived by discovering its
interesting properties when frustrated randomness is introduced. If we
assume that the exchange $J_{ij}$ is a random variable we speak about
a Potts glass. Its mean-field (long-range) version with a Gaussian
distribution $P(J_{ij})=(2\pi
J^{2}/N)^{-1/2}\exp\{-(J_{ij}-J_{0}/N)^{2}/2J^{2}/N\}$ is a
generalization of the Sherrington-Kirkpatrick (SK) model of a spin
glass \cite{Sherrington75}. Analogously to the pure model, the Potts
glass shows a richer critical behavior than the Ising spin glass
($p=2$) \cite{Elderfield83, Erzan83}. The peculiarity of the
mean-field solution of the p-state Potts glass for $p>2$ is that the
replica symmetric (RS) solution below the transition point to the
glassy phase does not break into the Parisi solution with a continuous
replica symmetry-breaking but rather to a first RSB hierarchical state
\cite{Gross85, Cwilich89}. The existence of a locally stable 1RSB
solution generated from an instability of the replica-symmetric one
may enlighten the construction of the the equilibrium glassy phase. In
the effort to solve the SK model, Parisi proposed a hierarchical
construction how to break the replica symmetry when using the replica
trick to average over the random spin couplings \cite{Parisi80}.
Although it is always necessary to construct the equilibrium glassy
state via discrete (integer) numbers of replicas, solutions with
finite many replica hierarchies were for long assumed as auxiliary,
intermediate states. Only the continuous Parisi solution was
considered to represent the true equilibrium. This was the case in the
SK model, where all $K$RSB solutions proved unstable. Local stability
of the 1RSB solution of the Potts glass renewed interest in solutions
with finite many hierarchies of replica generations.

The SK model proved very inspiring in understanding the
low-temperature glassy phase. Not only the replica-symmetry breaking
of Parisi was derived within it but also a mathematical poof was
presented that the construction of Parisi leads in this model to an
exact equilibrium state \cite{Guerra03,Talagrand06}. However, the
Ising spin glass has not allowed for understaning of the actual
meaning of the states with finite many hierarchies of replica
generations. The rigorous construction of Guerra and Talagrand does
not specify how many replica hierarchies are necessary to reach the
maximum of free energy. On the other hand, the $p$-state Potts model
of a glass can be used to clarify this point as well as the way how
and when, if ever, $K$RSB solutions break into the Parisi FRSB one. It
is by now clear that the 1RSB solution of the Potts glass does not
stay stable at all temperatures \cite{Gross85,Gribova03}. It breaks
down at a lower temperature $T_{G}$ to another state as discussed in
the framework of the $p$-spin glass \cite{Gardner85}. It is generally
assumed that the equilibrium state below this temperature is the
Parisi FRSB solution, though not rigorously
proven. Namely, it remains unclear when and how the FRSB solution
emerges in the $p$-state Potts glass or the $p$-spin glass.

It is the aim of this Letter to trace the genesis of the continuous
replica-symmetry breaking solution of the mean-field $p$-state Potts
glass.  We show that it emerges at the transition point to the glassy
phase and coexists with the 1RSB solution. Although the former is only
marginally stable and the latter locally stable, the FRSB
solution has a higher free energy and unlike 1RSB it is globally
\textit{thermodynamically homogeneous}. We demonstrate by explicit
calculations near the transition point to the glassy phase that 1RSB
is globally unstable and FRSB offers the true equilibrium with the
extremal free energy when the replica-symmetric solution becomes
unstable.

Thermodynamic homogeneity was introduced in spin glass models by one
of us \cite{Janis05} as a condition to be fulfilled by free energy
in the thermodynamic limit. Instability of the replica-symmetric
solution and replica-symmetry breaking in the replica trick were shown
to be equivalent to a breakdown of thermodynamic homogeneity. Thereby
a physical interpretation was given to replica-symmetry breaking. When
the statistical system is thermodynamically inhomogeneous the
thermodynamic limit of an isolated system (micro-canonical ensemble)
differs from the thermodynamic limit of a system embedded in a thermal
bath (canonical ensemble). The thermal bath can be generated by
scaling or replicating the original phase space. We probe a dependence
of the original system on the bath variables via an infinitesimal
interaction with the replicated ones.  Thermodynamically homogeneous
systems must not be macroscopically affected by such perturbations in the
thermodynamic limit. If the original system is macroscopically
influenced by an infinitesimal interaction with the bath, new
thermodynamic variables (chemical potentials) must be introduced to
equilibrate the system and the bath \cite{Janis06a}. Local
thermodynamic homogeneity guarantees only independence with respect to
small (infinitesimal) scalings of the phase space, that is, situations
with the bath in a thermal equilibrium. We show that the locally
stable 1RSB solution appears globally unstable and depends on the
variables of higher RSB solutions if the higher spin hierarchies
(bath) are pulled out of equilibrium. Only global thermodynamic
homogeneity, satisfied by FRSB, warrants that the equilibrium state
does not depend on the behavior of the bath (scaled or replicated
variables) in which the thermodynamic system is embedded, or with
which it can interact.

The starting point for the construction of the equilibrium state for
spin-glass systems is a free energy with $K$ spin hierarchies. We will
deal with the mean-field (long-range) $p$-state Potts model with
Hamiltonian $H_{P}$ from Eq.~\eqref{eq:Hamiltonian} and neglect all
spatial fluctuations. We assume a Gaussian randomness in the exchange
parameters $J_{ij}$ with $J_{0} = 0$ for $p\leq 4$ so that the
transition to the glassy phase is higher than that to a ferromagnetic
state \cite{Elderfield83}. A hierarchical free energy with $K$
hierarchies depends on $2K + 1$ variational parameters. It is a
parameter $q$ weighing the fluctuations of the random exchange and $K$
pairs $\Delta\chi_{l},m_{l}$ governing the interaction between the
hierarchies of replicated spin variables \cite{Janis05}. Mean-field
free energy with $K$ hierarchies for the $p$-state Potts glass has the
following form %
\begin{subequations}\label{eq:FE-K} 
  \begin{multline}\label{eq:freeE} -\beta f_{K}(q,\{\Delta\chi_{l}\},
    \{ m_{l}\}) =\frac{\beta^{2}J^{2}}{4}(p-1)\bigg\{\bigg(1-q \\ -
    \sum_{j=1}^{K}\Delta \chi_{j}\Bigg)^{2} -\sum_{j=1}^{K}m_{j}\Delta
    \chi_{j}\bigg[\Delta \chi_{j}\\ +2\bigg(q+\sum_{l=j+1}^{K}\Delta
    \chi_{l}\bigg)\bigg]\Bigg\} +\int \mathcal{D}_p y
    \log{z_{K}^{K}}(\mathbf{y}) \end{multline}
  where%
  \begin{equation}\label{eq:zK} z_{l}^{K}(\mathbf{y}) =\left[\int
      \mathcal{D}_p
      \lambda_{l}\left(z_{l-1}^{K})(\{\mathbf{\lambda}\}_l,\mathbf{y}\right)^{m_{l}}
\right]^{\frac{1}{m_{l}}}\   . 
\end{equation}%
The initial zero-level partial sum reads
  \begin{equation}\label{eq:zK} z_{0}\equiv
    z_{0}^{K}=\sum_{l=1}^{p}\exp\left\{\beta
      J\left(\sqrt{q}y^{\alpha}+\sum_{j=1}^{K}\sqrt{\Delta
          \chi_{j}}\lambda_{j}^{\alpha}\right)e^{\alpha}_{l}\right\}\ . 
  \end{equation}\end{subequations} %
We denoted a vector Gaussian measure $\mathcal{D}_p y = \prod_{\alpha
  =1}^{p-1}d y_{\alpha} \exp\left\{ -  \left(y^\alpha\right)^2/2\right\} /\sqrt{2\pi}$ and 
$\{\mathbf{\lambda}\}_l = \{\lambda_{l},\lambda_{l+1},\ldots
\lambda_{K}\}$. Notice that each integration variable $y, \lambda_{l}$
is a $p-1$ dimensional vector. It is worth noting that partition sum
$z_{0}$ depends on a particular representation of the basis vectors
$\mathbf{e}_{l}$, but the physical quantities, averaged over
fluctuating vector fields $\mathbf{y}$ and $\lambda_{l}$, are
invariant with respect to gauge transformations in the representation
space of the Potts spin-like variables.

The rigorous construction of Guerra and Talagrand tells us that the
equilibrium state is reached by supremum of functionals $f_{K}$ from
Eq.~\eqref{eq:FE-K} among all possible choices of parameters $K;q,
\Delta\chi_{l},m_{l}$. It, however, does not say whether the number of
hierarchies is finite or must be infinite. The Parisi continuous
symmetry-breaking solution within this construction is obtained as an
asymptotic limit $K\to\infty$ with an assumption of a homogeneous
distribution of overlap susceptibilities $\Delta\chi_{l}\to 1/K$
\cite{Janis06b}. Each hierarchical free energy $f_{K}$ satisfying $K$
stability conditions represents a local maximum if $0\le m_{l}\le 1$
\cite{Janis05}. The question is whether a locally stable solution with
a finite number of hierarchies is also the global maximum.
   
One of us recently derived a representation of the Parisi free energy
with FRSB in closed form being decoupled from the iterative
construction via discrete replica-symmetry breaking hierarchies
\cite{Janis08}. This representation allows us to look for a solution
with FRSB without investigating local stability of the discrete $K$RSB
solutions. It is a straightforward task to generalize the
representation from Ref.~\onlinecite{Janis08} to the $p$-state Potts
glass. We obtain %
\begin{subequations}\label{eq:FE-continuous}
  \begin{multline}\label{eq:freeEcontinuous} -\beta
    f_{c}[q,X,m(\lambda)]=\log p+\frac{\beta ^{2}}{4}(p-1)(1-q-X)^{2} \\
    -\frac{\beta ^{2}}{2}(p-1)X \int_{0}^{1}d\lambda
    m(\lambda)[q+X(1-\lambda)]+\left\langle g(1,\mathbf{h}+\mathbf{y}
      \sqrt{q})\right\rangle_{\mathbf{y}} \end{multline} %
  where %
  \begin{multline}\label{eq:E0} g(\nu,\textbf{h})
    =\mathbb{T}_{\lambda} \exp\bigg\{\frac{X}{2}\int_0^{\nu} d\lambda
    \left[
      \partial_{\bar{h}^{\alpha}}\partial_{\bar{h}^{\alpha}}
    \right. \\ \left.+ m(\lambda) g^{\prime}_{\alpha}(\lambda,
      \mathbf{h} + \bar{\mathbf{h}})\partial_{\bar{h}^{\alpha}}
    \right] \bigg\} g_{0}(\mathbf{h}+
    \mathbf{\bar{h}})\big|_{\bar{\mathbf{h}} = 0}
  \end{multline}\end{subequations}
and $ g_{0}(\mathbf{h})=\ln \sum_{l=1}^{p}\exp\{\beta
h^{\alpha}e^{\alpha}_{l}\}$ is the local interacting part of the Potts
free energy. We introduced an evolution operator represented via a
``time-ordering'' operator $\mathbb T_{\lambda}$ ordering products of
$\lambda$-dependent non-commuting operators from left to right in a
$\lambda$-decreasing succession. We further denoted
$g^{\prime}_{\alpha}(\lambda, \mathbf{h})
\equiv \partial_{h_{\alpha}}g (\lambda, \mathbf{h}) $.

Notice that the FRSB state is represented only by a single function,
$m(\lambda)$, $\lambda\in [0,1] $ in our case. It generalizes the
sequence $m_{l}$ from the discrete scheme. It is related to the Parisi
order-parameter function via $m(\lambda) = x(q + X(1 - \lambda))$. The overlap susceptibilities $\Delta\chi_{l}$ transform to $Xd\lambda$. The
Edwards-Anderson parameter here is $q_{EA}= q + X$. That is, our
function $m(\lambda)$ is essentially the inverse Parisi
order-parameter function (with the inverse slope).

We now have two free energies, $f_{K}$ from Eq.~\eqref{eq:FE-K} and
$f_{c}$ from Eq.~\eqref{eq:FE-continuous}, for which we can try to
find stationary solutions in the space of their variational
parameters. It is clear that we are able to resolve these free
energies and the stationary order parameters only near the continuous
transition to the glassy phase. To find the transition point we start
with the RS solution with the only parameter $q$
\cite{Elderfield83}. If we denote the transition temperature $T_{c}$
and $\tau=(T_{c}-T)/T_{c}$ we obtain in the leading order of $\tau$ the
asymptotics $q_{RS}\doteq 4\tau/(6 - p)$. We chose the energy scale
$J=1$.  The stability condition for the RS solution is $\Lambda_{0}
=2\tau(2-p)/(6-p)$ \cite{Cwilich89}. The RS solution is unstable
($\Lambda_{0}<0$) in the leading order of $\tau$ for $p>2$. The RS
solution of the Ising spin glass ($p=2$) gets unstable only in
$\tau^{2}$ \cite{Janis06b}.

We asymptotically expand free energies of the 1RSB and FRSB solutions
in powers of the small parameter $\tau$. We will calculate only the
differences of the free energies to the paramagnetic one, being $\beta
f_{para}=-\beta^2/4(p - 1)- \log(p)$. It is a rather tedious task to
do so and the details of this calculation will be presented
elsewhere. We only note here that we used the program MATHEMATICA to
generate the expansion coefficients of the solutions for the order parameters.
We had to expand them up to the second order for 1RSB and to the third order for
FRSB so that to obtain the expansion of the free energies to the fifth
order where the two solutions differ.

For 1RSB we derived the following asymptotic free energy
\begin{widetext}
  \begin{equation}\label{f:1RSB:tau} \frac{\beta}{p-1} f_{1RSB}\doteq
    \frac{\tau ^3}{3 (4-p)} + \frac{(p (11 p-102)+204) \tau ^4}{12 (4-p)^3} 
    -\frac{(p (p ((18744-1103 p) p-120648)+325728)-317232) \tau ^5}{720
      (4-p)^5} \ . \end{equation}\end{widetext} 
The variational parameters for
1RSB in the leading asymptotic order are $q\doteq 0$ and $\Delta\chi \doteq
2\tau/(4-p)$, $m_{1} \doteq (p - 2)/2 + (36 - 12p -p^{2}) \tau/4(4-p)$. The
transition from the RS to the 1RSB solution becomes discontinuous for
$p>4$. Unlike the RS solution, the 1RSB solution becomes locally stable for
$p>p^{*}$ as can be deduced from the stability criterion explicitly
evaluated near the transition point 
\begin{equation} \Lambda_{1} = \frac{\tau^{2}}{6(4 - p)^{2}} \left(
    7p^{2} - 24 p + 12\right) \ .
\end{equation} 
The stability function $\Lambda_{1}$ becomes positive for $ p \ge
p^{*}\approx 2.82$.

The expansion of the FRSB solution below the transition point to the
glassy phase is more intricate due to a complicated structure of
representation in Eq.~\eqref{eq:FE-continuous}. We have to solve
asymptotically equations for the order-parameter function $m(\lambda)$
for each $\lambda\in [0, 1]$. We find that $q=0$ identically and $X
\doteq 2 \tau/(4 - p)$ recovers in the leading order $\Delta\chi$ from
1RSB. Finally the order-parameter function in the leading-order reads
\begin{multline}
  m(\lambda) \doteq \frac{p-2}2 + \frac 12 \left[(3 - 2p) p + 6\right]
  X \\ + \frac 14\left[ (7p - 24) p + 12\right]\lambda X \ .
\end{multline}
Expansion of the free energy with the appropriate asymptotic solutions
for the order parameters then is
\begin{multline}\label{eq:ftau}
  \frac{\beta}{p-1} f_{c}(\tau)\doteq \frac{1}{3 (4-p)} \tau ^3
  +\frac{(p (11 p-102)+204) }{12 (4-p)^3}\tau ^4 \\ +\frac{(p (p (p
    (16 p-265)+1686)-4532)+4408) }{10 (4-p)^5}\tau ^5\ .
\end{multline}

\begin{figure} \includegraphics[scale=1]{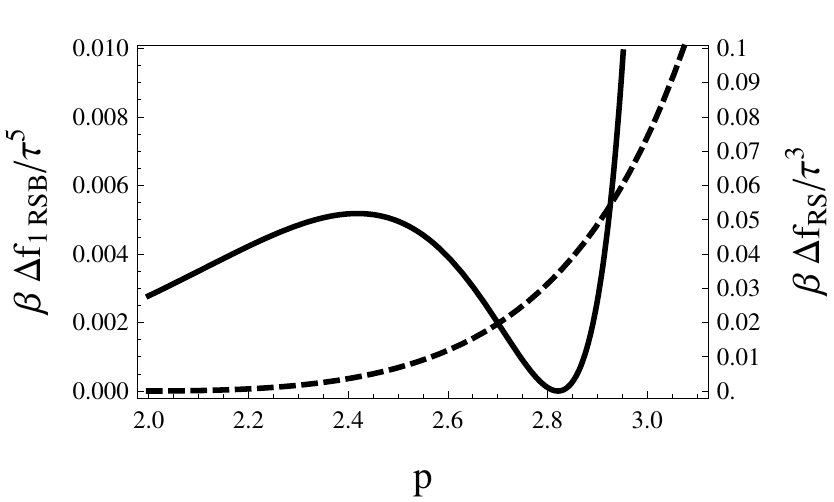} \caption{ Differences
    in the leading asymptotic order below the transition point to the
    glassy phase between the free energies of the FRSB and 1RSB (solid
    line) and FRSB and RS (dashed line) solution. Notice that the FRSB solution
    splits from the RS one at the order $\tau^{3}$ (right scale) and
    form 1RSB not till the fifth order $\tau^{5}$ (left
    scale).} \label{fig:dif-fe} \end{figure}
Comparing the asymptotic free energies for the 1RSB and FRSB solutions
we obtain a difference
\begin{equation}\label{eq:fdiff} \beta (f_{c}-f_{1RSB})\doteq
  \frac{(p-1) (p (7 p-24)+12)^2 \tau ^5}{720 (4-p)^5} 
\end{equation} 
telling us that the free energy from FRSB has for the Potts glass a
higher free energy than that from 1RSB for $ 2\le p \le 4$. Potts
glass for $p > 4$ demands a more detailed analysis, since the
denominator on the right-hand side of Eq.~\eqref{eq:fdiff} goes
through zero. We also calculated the difference of the FRSB solution
to the RS one leading to
\begin{equation}
  \beta(f_{c } - f_{RS}) \doteq \frac{(p - 1) (p - 2)^{2}\tau^{3}}{3(4 - p)(6
    - p)^{2}}\ . 
\end{equation} 
Notice that  $f_{c}= f_{1RSB}$ to this order. We see again that our
asymptotic expansion is meaningful only for $p\le 4$ where the
transition to the RSB phase is continuous and $\tau$ serves as a small
parameter. The $4$-state Potts glass is a singular (degenerate) case,
since $m_{1} =1$ at $T_{c}$ of the 1RSB solution. Both differences in
free energies are plotted as a function of $p$ in
Fig.~\ref{fig:dif-fe}. It is interesting to notice that $f_{1} \doteq
f_{c}$ for $p= p^{*}$ above which the 1RSB solution becomes locally
stable. The free energies equal, however, at this point only in the
leading order and this degeneracy is lifted in a higher order of
$\tau$.

Last but not least, the FRSB solution of the Potts glass suggests that
the physical interpretation of the Parisi order-parameter function
$q(x)$ should be reconsidered. Parisi \cite{Parisi83} proposed that
the derivative $dx(q)/dq$ be interpreted as the average of the
probability distribution $P(q)$ of different pure states with overlap
$q$. This interpretation seems not to work for the Potts glass,
since %
\begin{align}\label{eq:Overlap-distribution} P\left(q \right) &=-
  \frac{dm(\lambda)}{X d\lambda} \doteq \frac {p-2}{2X}\delta(\lambda)
  + \frac 14 \left[(24 - 7p)p - 12 \right] \end{align}
where $q = X(1 - \lambda)$. Function $P(q)$ gets negative for $p>p^{*}
\approx 2.82$, the stability region of the 1RSB solution.

To conclude, our analysis of the asymptotic solution of the $p$-state
Potts glass for $2 \le p \le 4$ below the continuous transition to the
glassy phase revealed, against general expectations, that the 1RSB
solution is not the true equilibrium, although locally stable for
$4\ge p \ge p^{*}\approx 2.82$. We found a different state with FRSB
that continuously develops from the critical point $T_{c}$. The FRSB
solution has a higher free energy than that of the 1RSB one and
represents the global maximum as demanded by the rigorous construction
of Guerra and Talagrand. Below the critical point $T<T_{c}$ for $2\le p \le 4$, the 1RSB solution is globally unstable and decays on a long but finite time scale to the FRSB
solution. We could give the derived result a physically appealing
interpretation in that the 1RSB solution is only locally
thermodynamically homogeneous while the necessary condition of global
thermodynamic homogeneity is fulfilled only by the solution with the continuous
FRSB. 

Research on this problem was carried out within project AV0Z10100520
of the Academy of Sciences of the Czech Republic.

 \end{document}